\documentclass[11pt]{article}
\usepackage{eurosym}
\usepackage{amsmath}
\usepackage{amsfonts}
\usepackage{amssymb}

\begin{document}

\title{\textbf{$\theta$-term in a bounded region}}
\author{\textbf{Fabrizio Canfora}$^{1}$, \textbf{Luigi Rosa}$^{2}$ and 
\textbf{Jorge Zanelli}$^{1}$ \\
$^{1}$Centro de Estudios Cient\'{\i}ficos (CECs), Casilla 1469, Valdivia,
Chile.\\
$^{2}$Dipartimento di Fisica, Universita' Federico II Napoli, Italy.}
\maketitle

\begin{abstract}
We analyse the physical implications of adding a topological density term $\theta Tr(F\wedge F)$ to a gauge theory in a bounded region. In particular, we calculate the Casimir effect on a spherical region and we show that the result is not periodic in $\theta $, contrary to what would be expected for a true topological density. The topological nature of the $\theta$-term can be restored if an additional boundary term required by the Atiyah-Patodi-Singer theorem is included. Then, the periodicity is trivially restored because the resulting Casimir energy is independent of $\theta$.
The results of the present work suggest that \textit{the observable effects of the} $\theta$ \textit{term could be very small even without assuming} $\theta$ \textit{itself to be small}.
\end{abstract}

e-mails: {\small canfora-at-cecs.cl} {\small rosa-at-na.infn.it} {\small %
z-at-cecs.cl}

%%%%%%%%%%%%%%%%%%%%%%%%%%%%%%%
\section{Introduction}
%%%%%%%%%%%%%%%%%%%%%%%%%%%%%%%

The existence of instantons, one of the most remarkable non-perturbative effects in QCD, led to the resolution of the $U(1)$ problem \cite{tHooft}: 't Hooft's key observation was that, due to the presence of anomalies \cite{anom1} \cite{anom2} \cite{anom3}, instantons can solve this problem provided the term $\theta Tr(F\wedge F)$ is added to the QCD action.\footnote{This is sometimes written as $\frac{\theta }{16\pi ^{2}}\epsilon ^{\mu \nu \lambda \rho }Tr(F_{\mu \nu }F_{\lambda \rho })$, but here we will absorb the normalization in $\theta$ and use exterior forms whenever possible.} (a pedagogical discussion of the problem can be found in \cite{Coleman}) This term is proportional to the instanton number density, a topological quantity that does not affect the field equations or the pertubative expansion around the vacuum. The addition of this term to the Euclidean action changes the weight in the path integral for configurations with different instanton number, and this opens a nonperturbative channel for parity-violating decays. Another important feature of this term is that since the instanton number takes only integer values, quantum mechanics requires periodicity of the observable quantities under $\theta \rightarrow \theta +2\pi n$, for
some fixed integer $n$.

Despite the elegance and effectiveness of 't Hooft's argument, it raises another question, namely, why other observable effects of the $\theta $-term are so small. For instance, the experimental data on the electric dipole moment of the neutron imply that if the corrections \textit{are proportional to} $\theta$ \textit{itself}, then $|\theta |<10^{-9}$ \cite{Crewther:1979pi}. This so-called strong CP problem has no satisfactory explanation up to now (for detailed reviews, see \cite{Vicari,KimCP,Peccei,Posp}).

The aim of this note is to analyse the effect of the $\theta$-term in a bounded region such that the value of $\theta$ jumps across the boundary of the region itself. In this example, we take a spherical region, where $\theta=\theta_{in}$, surrounded by a medium with $\theta=\theta_{out}$. In a sense, this is the simplest possible scenario \textit{a la} Peccei-Quinn \cite{PecceiQuinn}, where $\theta$ is not a dynamical field, but takes different constant values in different domains. The situation we want to describe is one in which the vacuum expectation value of the Peccei-Quinn field in the deconfined phase can be different from its value in the confining phase\footnote{Similar ideas are quite common in cosmology (see the classic paper \cite{cosmobubble}; a modern treatment can be found in \cite{cosmobubble2}) where it is interesting to study bubble of vacua evolving inside a cosmological fluid belonging to a different vacuum (the different vacua being distinguished, for instance, by the value of the cosmological constant). In the present case, the bounded region is distinguished from the "exterior" region by the value of $\theta$.}. 

The presence of physical boundaries cannot be ignored in many situations. For instance, the external surface of a hadron in QCD acts as a boundary separating a region where gluons and quarks propagate freely, from the outer region in which there are no free quarks or gluons. This is the basic hypothesis of the well-known MIT bag model (for a review, see \cite{Mitmodel}). Also in gravitation theories, the horizon of a black hole can also be
thought of as an effective boundary, responsible for the Hawking radiation.

Although the addition of a topological density does not affect the classical field equations it does induce new effects if the region where the fields are defined is bounded. In particular, in order to make the action principle well defined, the boundary conditions must be changed. This in turn reflects on the dynamics of the fields in the bulk by modifying the spectrum of the theory even at the linearized (perturbative) level.

There are at least two alternative ways of interpreting the presence of a $\theta $-term in the action in a bounded region. One can assume the boundary conditions appropriate for that action, or one can modify the action by the addition of a boundary term so that the $\theta $-term continues to represent the instanton number of the original proposal by 't Hooft.

Here we compute the Casimir energy in a cavity and examine these two possibilities: i) the action, with the $\theta $-term is integrated over the bounded region, assuming boundary conditions consistent with the variational principle; ii) the action is supplemented by the addition of a boundary term so that the $\theta$-term remains a topological correction. In the first case, the boundary conditions acquire a twist proportional to $\theta$ and the resulting Casimir energy picks up a correction quadratic in $\theta$ and is not invariant under\footnote{In natural units, $\hbar=c=1$.} $\theta \rightarrow \theta +2\pi n$ for any
value of $n$. In the second case, it is easily seen that there can be no perturbative effects of any kind, and the Casimir energy is insensitive to the presence of a $\theta$-term.

We thus conclude that the observable effects of $\theta$ should not be expected to be \textit{proportional to} $\theta $ \textit{itself}. In one interpretation, the corrections to the Casimir energy are neither proportional to, nor periodic in $\theta $. In the other, there is no need to assume $|\theta |$ to be extremely small in order to explain the lack of experimental evidence for parity violation of perturbative nature in QCD, as the effects of the $\theta$-term would cancel out to all orders in the perturbative expansion.

The paper is organized as follows. In section 2, the variational principle in the presence of a $\theta$-term in a bounded region is discussed. Section 3 examines how the presence of a $\theta$-term affects the Casimir energy. Section 4 discusses how the topological nature of the $\theta$-term could be restored by adding a boundary term. Finally, section 5 contains the conclusions, and the details of the calculation of the Casimir energy can be
found in the appendix.

%%%%%%%%%%%%%%%%%%%%%%%%%%%%%%%%%%%%%%%%%%%%%%%%%%%
\section{Boundary conditions induced by the $\protect\theta$-term}
%%%%%%%%%%%%%%%%%%%%%%%%%%%%%%%%%%%%%%%%%%%%%%%%%%%

In order to examine the physical effects of a $\theta $-term in the presence of boundaries, let us consider the Abelian case, for the sake of simplicity. Without loss of generality, here we will assume that $\theta_{in}\neq 0$ while $\theta_{out}=0$, so that the jump $\theta _{in} - \theta_{out}$ is simply $\theta$. The generalization to a non-Abelian gauge theory could be easily discussed along the same lines. Thus, we are analyzing the case in which the electromagnetic field with a $\theta $-term is confined in a region $M_{in}$ in spacetime, the total spacetime being $M=M_{in}\cup M_{out}$. The corresponding action is\footnote{The two-form $F=\frac{1}{2}F_{\mu \nu }dx^{\mu }\wedge dx^{\nu }=dA$ is the ordinary electromagnetic field, and $A$ is the $U(1)$ connection 1-form. The Hodge dual is defined as $\ast F_{\mu \nu }=\frac{1}{2}\epsilon _{\mu \nu\lambda \rho }F^{\lambda \rho }$.} 
\begin{equation}
I[A]=\frac{1}{2}\int_{M}\left( F\wedge \ast F\right) -\frac{\theta }{2}\int_{M_{in}}\left( F\wedge F\right) \ .  \label{(0)}
\end{equation}
In the usual tensor notation, the action reads 
\begin{equation}
I[A]=\frac{1}{2}\int_{M}\left( F_{\mu \nu }F^{\mu \nu }\right) -\theta \frac{1}{2} \int_{M_{in}} \left(\epsilon^{\mu \nu \rho \sigma} F_{\mu \nu}F_{\rho \sigma}\right) d^{4}x.  \label{usual1}
\end{equation}
Here $M_{in}=D\times \mathbb{R}$, where $D$ is a three-dimensional spatial ball with boundary $\partial D=S^{2}$, and $\mathbb{R}$ is the time axis. The boundary is $\Sigma =\partial M_{in}=S^{2}\times \mathbb{R}$.

Varying the action with respect to $A_{\nu}$ yields 
\begin{equation}
\delta I[A]=-\int_{M}\delta A_{\nu}\left( \partial _{\mu}F^{\mu \nu}\right) d^{4}x + \int_{\Sigma}n_{\mu}\delta A_{\nu} \left( F^{\mu \nu} +\theta \epsilon^{\mu \nu \rho \sigma}F_{\rho \sigma}\right) d^{3}x\ ,
\label{varprinc1}
\end{equation}
where $n_{\mu }$ is the outward normal to $\Sigma $. As usual, requiring the first term on the right hand side to vanish for an arbitrary variation $\delta A$ implies the Maxwell equations in $M$, $\partial_{\mu} F^{\mu\nu}=0 $. The second term provides a recipe to assign boundary conditions consistent with the requirement that $I[A]$ should have an extremum when the classical field equations are satisfied in the bulk (see e. g., \cite{HRT,RT}), 
\begin{equation}
\delta A_{\mu}n_{\nu}\left( F^{\mu \nu}+\theta \epsilon^{\mu \nu \rho \sigma}F_{\rho \sigma}\right) |_{\Sigma}=0.  \label{varprinc2}
\end{equation}

We see that although the $\theta $-term is locally a total derivative --and hence it does not modify the field equations--, it requires a change in the boundary conditions if the variational principle is to be well defined. The $\theta$-term produces a duality rotation of the components of the field strength along the normal of $n_{\mu}$ at the boundary. In other words, consistency of the variational principle requires replacing the standard Neumann boundary conditions on $\left. n_{\nu}F^{\mu \nu}\right\vert_{\Sigma}=0$, by the modified expression 
\begin{equation}
\left. n_{\nu}\left( F^{\mu \nu} + \theta \epsilon^{\mu \nu \rho \sigma} F_{\rho \sigma}\right) \right\vert_{\Sigma}=0 \,.  \label{replaceingdata}
\end{equation}
Note that (\ref{varprinc2}) is the only boundary condition consistent with path integral quantization: when the classical field equations are satisfied the action \textit{must be stationary}, otherwise the Feynman interpretation would not work.

An equivalent form to state the same result is that the action (\ref{usual1}) gives rise to a set of Maxwell equations with a current source supported at the boundary, 
\begin{equation*}
d\ast F=j\wedge F\text{ \ ,}
\end{equation*}
or, in more familiar notation, 
\begin{equation}
\partial_{\mu}F^{\mu \alpha}=\frac{\theta }{2}\delta (\Sigma)\epsilon^{n\alpha \mu \nu}F_{\mu \nu}\, ,  \label{3}
\end{equation}
where the index $n$ refers to the normal direction to $\Sigma$. Thus, the effect of the $\theta$-term is to introduce an effective current density on the surface of the region $M$. The peculiar feature of this ``source" term is that it is proportional to the components of the electromagnetic field itself, and therefore the superposition principle (linearity of the
equations) still holds.

The equivalence between (\ref{3}) and the standard Maxwell equations with the modified boundary conditions (\ref{replaceingdata}) can be understood from the fact that since the $\theta$-term is \textit{locally} a total derivative, it can be replaced by a boundary term that represents the presence of a source at the boundary \cite{Huerta-Zanelli}.

Writing (\ref{3}) in coordinates adapted to the surface, one finds\footnote{Here $(\overrightarrow{E})_{i}=F^{0i}=-F_{0i}$ and $(\overrightarrow{B})i=\frac{1}{2}\epsilon _{ijk}F^{jk}$.} 
\begin{eqnarray}
\mathbf{\nabla }\cdot \mathbf{E} &\mathbf{=}&\theta \delta (\Sigma )\mathbf{B}\cdot \mathbf{n}  \label{Gauss} \\
-\partial_{t}\mathbf{E+\nabla \times B} &\mathbf{=}&\theta \delta (\Sigma) \mathbf{E\times n}  \label{Ampere}
\end{eqnarray}
where $\mathbf{n}$ is the unit normal to $\Sigma$. Assuming that the time derivatives of the fields are finite \cite{jack62}, in the vicinity of the surface $\Sigma$ these equations imply that the normal component of $\mathbf{E}$, and the tangential components of $\mathbf{B}$, are discontinuous, 
\begin{eqnarray}
\lbrack \mathbf{E}_{n}] &=&\theta \mathbf{B}_{n}  \label{4} \\
\lbrack \mathbf{B}_{\Vert }] &=&-\theta \mathbf{E}_{\Vert }  \label{5}
\end{eqnarray}

On the other hand, from the Bianchi-Jacobi identity $dF\equiv 0$ ($\mathbf{\nabla }\cdot \mathbf{B}=0$, $\partial _{t} \mathbf{B+\nabla \times E}=0$), it follows that the normal component of $\mathbf{B}$ and the tangential component of $\mathbf{E}$ must be continuous (in the static case at least), 
\begin{eqnarray}
\left[ \mathbf{B}_{n}\right] &=&0  \label{6} \\
\left[ \mathbf{E}_{\Vert }\right] &=&0  \label{7}
\end{eqnarray}
These continuity conditions imply that the right hand sides of (\ref{4}) and (\ref{5}) are well defined and they represent surface charge and current densities, respectively.

The phenomenological novelty here is that these sources are given by components of the electromagnetic field itself. The ``surface charge" in (\ref{Gauss}) is proportional to the normal component of the magnetic field, which is similar to the behaviour of vortices with magnetic flux as carriers of electric charge in superconductors. On the other hand, the tangential component of the electric field plays the role of a ``surface current" in (\ref{Ampere}), this surface current makes the region with non vanishing $\theta$ to behave like a topological insulator, a material that possesses surface current but is an insulator in the bulk \cite{Hassan-Kane}.

%%%%%%%%%%%%%%%%%%%%%%%%%%%%%%%%%%%%%%%%
\section{Casimir effect with a $\protect\theta $-term}
%%%%%%%%%%%%%%%%%%%%%%%%%%%%%%%%%%%%%%%%

We now examine how the Casimir energy is modified by the presence of the $\theta $ term through the replacement of Eq.(\ref{replaceingdata}) in an Abelian gauge theory (the computation in the non-Abelian case is analogous \cite{brevik82,plunien86}). Let us consider the case in which $D$ is a three-dimensional sphere. Details of the calculation can be found in appendix A (where the QCD case is also considered), quoting in this section only the relevant formulas. The regularized Casimir energy can be written as 
\begin{equation}
E_{reg}=\frac{1}{\pi R}\sum_{l=1}^{\infty }(l+1/2)\int_{0}^{\infty }{dy\ln [1-\eta \lambda _{\nu }^{2}(y)]}  
\label{casi2}
\end{equation}
with $R$ the radius of the sphere, where $\nu =l+1/2$, 
\begin{equation}
\eta =\frac{4\xi ^{2}\epsilon _{1}^{2}+\theta ^{2}(1+\xi )^{2}}{4\epsilon_{1}^{2}+\theta ^{2}(1+\xi )^{2}}=\frac{(\epsilon
_{1}-\epsilon_{2})^{2}+\theta ^{2}}{(\epsilon _{1}+\epsilon _{2})^{2}+\theta^{2}}\leq 1,\ \ \xi =\frac{(\epsilon _{1}-\epsilon _{2})}{(\epsilon_{1}+\epsilon _{2}) }\ \ ,  \label{def1}
\end{equation}
and $\epsilon _{1}$, $\epsilon _{2}$ are the dielectric constant inside and outside the sphere, respectively. We note that for $\theta =0$ the standard result is recovered \cite{brene98}: 
\begin{equation}
E_{reg}=\frac{1}{\pi R}\sum_{l=1}^{\infty }(l+1/2)\int {dy\ln [1-\xi^{2}\lambda _{\nu }^{2}(y)]}\,.
\end{equation}
Also, in the case of a perfect conducting sphere, $\xi ^{2}\rightarrow 1$, the contribution from the $\theta $-term disappears and the usual result is obtained \cite{badu78,milton78,nespi98}, 
\begin{equation}
E_{reg}=\frac{1}{\pi R}\sum_{l=1}^{\infty }(l+1/2)\int {\ dy\ln [1-\lambda_{\nu }^{2}(y)]}\,.  \label{eq:ereg}
\end{equation}
This is because for a perfect conductor the tangential component $\mathbf{E}_{||}$ vanishes identically, so the boundary conditions are not affected if $\theta \neq 0$. Note that the $\theta $ contribution simply reduces to a change of the $\xi $ parameter, $\xi ^{2}\rightarrow \eta $. The integral is convergent see \cite{nespi98}, but the $\nu $ series needs to be regularized. Using $\zeta $ function techniques, we find (see appendix), 
\begin{equation}
E=\frac{3\eta }{64R}-\frac{27\eta }{16384R}(\pi ^{2}-8)+\frac{63\eta ^{2}}{32768R}(\pi ^{2}-8).  \label{eq:casie}
\end{equation}
Finally, restoring the constants $\hbar $ and $c$, the Casimir Energy is 
\begin{equation}
E=\frac{\hbar c}{2R}(0.099912\eta -0.007189\eta ^{2}),
\end{equation}
and in the limit $\xi \rightarrow 1$, $\eta =1$ we obtain, at this order in $\nu $, 
\begin{equation}
E=\frac{0.092723}{2R}\,,  \label{E(R)}
\end{equation}
in very good agreement with earlier results \cite{badu78,milton78,nespi98}.

Note that, in principle, the Casimir effect provides a way to detect the presence of a jump in $\theta $.

It is worth noting that in the present framework there are two kinds of discontinuities: one is the usual jump in the dielectric constant at the interface between two materials. The other reflects the possibility of having different values of theta in different regions, for instance $\theta \neq 0$ inside and $\theta =0$ outside \cite{borvas00, gruco11}. This situation is a simplified version of the Peccei-Quinn mechanism in which $\theta $ is a dynamical field which should be varied in the action principle.

For $\theta =\theta _{in}-\theta _{out}=0$ the Casimir energy vanishes if the dielectric constants inside and outside the sphere are equal, $\epsilon_{1}=\epsilon _{2}$, and $\eta =\xi ^{2}=0$. However, for $\theta =\theta_{in}-\theta _{out}\neq 0$, $\eta $ is non-vanishing even in that case, and for $|\epsilon _{1}-\epsilon _{2}|\ll 1$, the Casimir energy is sensitive to the presence of a $\theta $ term. On the other hand, if $\theta$ is very large, there is a sort of \textquotedblleft universal" behaviour since $\eta $ approaches $1$ independently of the values of $\epsilon _{1}$ and $\epsilon _{2}$: 
\begin{equation}
\eta \underset{\theta \rightarrow \infty }{\rightarrow }1\ .  \label{eta}
\end{equation}

Thus, the addition of the topological term in (\ref{(0)}) can change the phenomenology of the Casimir effect in an observable manner. For similar conclusions in different context see also \cite{borvas00, gruco11}. This result shows, in particular, that the Casimir energy (\ref{eq:casie}) depends on $\theta $ in a non-trivial way through $\eta $, and that \textit{it is not periodic in} $\theta $. As it will be discussed in the Appendix, assuming the correcteness of the hypothesis behind the MIT bag model, the results in the case of a non-Abelian gauge theory would be analogous to the expression in Eq. (\ref{eq:casie}) so that also in the non-Abelian case the Casimir energy is not a periodic function of the jump in $\theta$. Such lack of periodicity may indicate that a jump in $\theta $ across the surface of the bag would not produce a truly topological effect. In the next section, it will be shown how to deal with this issue.

%%%%%%%%%%%%%%%%%%%%%%%%%%%%%%%%%%%%%%%%%%%%%%
\section{$\protect\theta$-term in the presence of boundaries}
%%%%%%%%%%%%%%%%%%%%%%%%%%%%%%%%%%%%%%%%%%%%%%

In the case of a compact manifold $M$ without boundary, the periodicity in $\theta $ of the path integral is expected on topological grounds: the partition function of a non-Abelian gauge theory coupled to fermions is, 
\begin{equation}
Z[A,\psi ]=\int [dA][d\psi ]\exp {\ \frac{i}{\hbar }\left[ I_{0}+\theta \int_{M}Tr\left( F\wedge F\right) \right] }\ .
\end{equation}
The second term in the exponential is the Pontryagin number, a topological invariant whose value is quantized, 
\begin{equation}
\int_{M}Tr\left( F\wedge F\right) =n(M)\in \mathbb{Z}.
\label{Pontryagin}
\end{equation}
Thus, if $M$ is compact and without boundary, the path integral, as well as all observable physical quantities should be periodic in $\theta $ (see, for instance, \cite{Coleman}).

As explicitly seen in the Casimir energy formula (\ref{casi2}), this is not true if $M$ has a non-trivial boundary $\partial M =\Sigma$, so that $\Sigma$ is the surface of discountinuity of $\theta$. In that case, it makes sense to analyze the dependence of physical observables on $\theta =\theta _{in}-\theta _{out}$: it is natural to assume that inside the bag, where gluons are free $\theta_{in}\neq 0$, while outside the bag where the gluons are confined $\theta_{out}$ vanishes. The periodicity of the Casimir energy on $\theta $ can be achieved by the addition of a boundary term on $\Sigma =\partial M$ that cancels the one coming from the variation of the $\theta $-term. 

Indeed, the correct expression of the topological invariant that replaces (\ref{Pontryagin}) for a manifold with boundary takes the form \cite{Nakahara}
\begin{equation}
\int_{M}Tr(F\wedge F)-\left( \int_{\partial M}Tr(A\wedge dA+\frac{2}{3} A\wedge A\wedge A)+\eta (\partial M)\right) =n(M)\in \mathbb{Z}\,,
\label{Pontryagin*}
\end{equation}%
where the second term is the Chern-Simons form at the boundary. The last term, $\eta (\partial M)$, a spectral invariant given by the index of a Dirac operator defined intrinsically on $\partial M$, as in the Atiyah-Patodi-Singer theorem, is not relevant in the case of $M$ = $D\times \mathbb{R}$, where $D$ is a three-dimensional spatial ball (which can be considered as the ground state of the bag). Of course, in the cases of more complex topologies (which, from the point of view of the MIT bag model, represent highly excited states of the bag itself outside the domain of perturbation theory) the last term, $\eta (\partial M)$ does not vanish so that in the non-perturbative regime it gives rise to non-trivial contributions. Thus, the left hand side of Eq. (\ref{Pontryagin*}) is the index of the Dirac operator on a manifold with boundary. The invariance of (\ref{Pontryagin*}) under arbitrary continuous deformations of $A$ can be directly checked. In particular, the variation of the second term exactly cancels the extra boundary term in (\ref{varprinc1}).

Thus, in order for the $\theta $-term to be a true topological invariant in a manifold with boundary, the action (up to a constant) should be of the form 
\begin{equation}
I[A;M]=I_{0}+\left[ \theta \int_{M}Tr(F\wedge F) - \theta \int_{\partial M}Tr(A\wedge dA+\frac{2}{3}A\wedge A\wedge A)\right] \,,  \label{improved}
\end{equation}
where $I_{0}$ is a gauge-invariant action (like the fermionic part of the QCD action), $M$ = $D\times \mathbb{R}$. In this case, all perturbative quantities such as cross sections or correlation functions derived form the improved action would not depend on $\theta$. On the other hand, in the cases of topologies more complex than $D\times \mathbb{R}$ (representing highly excited states of the bag\footnote{It is worth noting that the string picture of QCD provided by the large N expansion (see \cite{T74a} \cite{T74b} \cite{Ve76} \cite{Wi79}) suggests that bag states with non-trivial topologies are suppressed in the topological expansion.}) the $\eta (\partial M)$ in Eq. (\ref{Pontryagin*}) would also give rise to non-trivial contributions.

%%%%%%%%%%%%%%%%%%%%%%%%%%%%%%%
\section{Summary and Conclusions}
%%%%%%%%%%%%%%%%%%%%%%%%%%%%%%%

We have computed, in the presence of a $\theta $-term in the electromagnetic action, the Casimir energy in a cavity allowing for the possibility of having different values of $\theta$ inside and outside the cavity, which can be interpreted as different vacua of the Peccei-Quinn field. The Casimir energy is not a periodic function of the jump in $\theta $ across the boundary of the cavity itself, which is due to the fact that the $\theta $-term is not a topological invariant in a bounded region. This can be seen in different equivalent ways: it may be attributed to the presence of an effective ``source" $\theta \ast F$ at the boundary, or equivalently, to the modification of the boundary conditions needed by consistency of the action principle, or more abstractly, to the need to modify the Pontryagin invariant by adding the boundary terms suggested by the Atiyah-Patodi-Singer theorem in order to render the $\theta $-term a true topological invariant.

Modifying the $\theta$-term as in (\ref{improved}), in order to restore periodicity, makes the new action incapable to produce any observable effects in a perturbative expansion. In particular, the new $\theta $-term, being locally a total derivative, does not contribute to the field equations, the propagators, or the vertices, and cannot contribute to the Casimir energy either (because it cancels out in the boundary conditions). It is truly a topological invariant.

With the modified action, $\theta$ could only show up through non-perturbative effects which, typically, behave as $\exp(-\theta/g)$ or $\exp \left(-\theta/g^2\right) $. The interesting conclusion is that the observable effects of the $\theta $ term could be extremely small even if $\theta $ itself is not. This could help to understand the smallness of the observable parity violating effects of $\theta $-term in QCD if the topological term (\ref{Pontryagin*}) is used, which would be the correct one
in the presence of boundaries.

The possibility of interpreting the term $\theta F\wedge F$ in a bounded region as a boundary effect offers an interesting way of generalizing a gauge theory by introducing a non-standard coupling between the connection and a membrane as discussed in \cite{Z,E-Z,M-Z,EGMZ}. In fact, one can think of the Chern-Simons ($2p+1$)-form as the natural way to couple a gauge theory to a $2p$-brane, which in the case at hand is the boundary $\partial M $ of the cavity where the Casimir energy was computed. These results manifest an interesting similarity with the Callan-Rubakov effect \cite{Callan-Rubakov} which would be worthy of further investigation.

%%%%%%%%%%%%%%%%%%
\section*{Appendix A}
%%%%%%%%%%%%%%%%%%

In the following we will concentrate on the electromagnetic field. Neglecting the contribution of the non-linear term $f_{abc}A_b^{\mu}A_c^{\nu} $ to the field-strenght tensor, the results we obtain are general enough to be applied to the QCD bag model \cite{brevik82,plunien86} under the usual constraint $\epsilon \mu =1$, so that the gluon velocity in the medium is 1. In such approach \emph{perfect color confinement} is obtained considering the vacuum exterior to the bag (which in the following we will indicate with the suffix 2) as a perfect color magnetic conductor: $\mu_{2}\rightarrow \infty $, $(\epsilon _{2}=0)$, while the vacuum inside the bag (indicated with the suffix 1) is characterized by a color magnetic permeability $\mu _{1}=1$, $(\epsilon_{1}=1)$ (see considerations after eq. (\ref{eq:ereg})). With these assumptions, the corresponding Casimir energy is obtained simply multiplying eq. (\ref{eq:ene}) by eight \cite{brevik82}, see also \cite{plunien86,milton80} and references therein.

Indeed, the extrapolation of the Abelian results for the Casimir energy to the non-Abelian case is reasonable only if the hypothesis behind the MIT bag-model can be trusted. Namely, we are assuming, as it is commonly done (see, for instance, \cite{brevik82,plunien86}), that inside the bag the non-linear terms can be neglected and the vacuum is perturbative while outside there is a perfect color magnetic conductor. Obviously, in the case in which the QCD vacuum cannot be described in this simplified way, one should compute the Casimir energy based on the real QCD vacuum. On the other hand, the real non-perturbative QCD vacuum is still not available, and we think that a computation of the $\theta$-dependent Casimir effect based on the hypothesis of the MIT bag-model is sufficiently interesting because of the many phenomenological applications of the MIT model itself.

Let us consider the case in which $D$ is a three-dimensional sphere. In the interior of the sphere the fields take the form 
\begin{eqnarray*}
\mathbf{E}^{1} &=&\sum_{l=1}^{\infty }\left\{ \frac{i}{rk_{1}}a_{TE}^{1} %
\left[ i\hat{\mathbf{n}}j_{\nu }(k_{1}r)Y_{lm}(\theta ,\phi
)+(k_{1}rj_{\nu}(k_{1}r))^{\prime }\hat{\mathbf{n}}\times \mathbf{X}_{lm}%
\right] \right. \\
&&\left. +a_{TM}^{1}j_{\nu }(k_{1}r)\mathbf{X}_{lm}\right\} \\
\mathbf{H}^{1} &=&\sum_{l=1}^{\infty }\frac{k_{1}}{\omega \mu _{1}}%
\left\{a_{TE}^{1}j_{\nu }(k_{1}r)\mathbf{X}_{lm}-i\frac{a_{TM}^{1}}{rk_{1}}%
\left[ i\hat{\mathbf{n}}j_{\nu }(k_{1}r)Y_{lm}(\theta ,\phi )\right. \right.
\\
&&\left. \left. +(k_{1}rj_{\nu }(k_{1}r))^{\prime }\hat{\mathbf{n}}\times 
\mathbf{X}_{lm}\right] \right\}
\end{eqnarray*}%
where $Y_{lm}$ and $\mathbf{X}_{lm}$ are the spherical harmonics and the
vectorial spherical harmonics respectively, $k_{i}=\sqrt{\epsilon_{i}\mu_{i}}%
\omega $ $i=(1,2)$, $j_{\nu }(x)=\sqrt{\frac{\pi }{2x}}J_{l+1/2}(x)$ are the
spherical Bessel functions \cite{jack62}, and $(xf(x))^{\prime }\equiv \frac{%
d}{dx}(xf(x))$. In the exterior, the fields can be obtained by the previous
equations by simply substituting $a^{1}\rightarrow
a^{2},j_{\nu}(x)\rightarrow h_{\nu }^{1}(x)$ (the Hankel function \cite%
{jack62}), $k_{1}\rightarrow k_{2},(\epsilon _{1},\mu _{1})\rightarrow
(\epsilon_{2},\mu _{2})$. By imposing the boundary conditions on the
tangential components, $[\mathbf{D}_{||}]=0;[\mathbf{H}_{||}]=-\theta 
\mathbf{E}_{||}|_{\Sigma }$, we obtain the following equations: 
\begin{eqnarray*}
\frac{1}{k_{1}}a_{TE}^{1}(k_{1}Rj_{\nu }(k_{1}R))^{\prime }-\frac{1}{k_{2}}
a_{TE}^{2}(k_{2}Rh_{\nu }^{1}(k_{2}R))^{\prime } &=&0 \\
-\frac{1}{\omega \mu _{1}}a_{TM}^{1}(k_{1}Rj_{\nu }(k_{1}R))^{\prime }+\frac{%
1}{\omega \mu _{2}}a_{TM}^{2}(k_{2}Rh_{\nu }^{1}(k_{2}R))^{\prime }-\frac{%
\theta }{k_{2}}a_{TE}^{2}(k_{2}Rh_{\nu }^{1}(k_{2}R))^{\prime } &=&0 \\
a_{TM}^{1}j_{\nu }(k_{1}R)-a_{TM}^{2}h_{\nu }^{1}(k_{2}R) &=&0 \\
\frac{k_{1}}{\omega \mu _{1}}a_{TE}^{1}j_{\nu }(k_{1}R)-\frac{k_{2}}{%
\omega\mu _{2}}a_{TE}^{2}h_{\nu }^{1}(k_{2}R)-\theta
a_{TM}^{2}h_{\nu}^1(k_{2}R) &=&0.
\end{eqnarray*}

The determinant of the system is 
\begin{eqnarray}
Det(\omega ,R) &=&\frac{1}{k_{1}k_{2}}\biggl\{\theta^{2}h_{%
\nu}^{1}(k_{2}R)j_{\nu }(k_{1}R)(k_{2}Rh_{\nu }^{1}(k_{2}R))^{\prime
}(k_{1}Rj_{\nu }(k_{1}R))^{\prime }+ \\
&&\frac{1}{(\mu _{1}\mu _{2}\omega )^{2}}\left[ \mu_{1}j_{%
\nu}(k_{1}R)(k_{2}Rh_{\nu }^{1}(k_{2}R))^{\prime }-\mu _{2}h_{\nu
}^{1}(k_{2}R)(k_{1}Rj_{\nu }(k_{1}R))^{\prime }\right] \times  \notag \\
&&\left[ \mu _{1}k_{2}^{2}h_{\nu}^{1}(k_{2}R)(k_{1}Rj_{\nu}(k_{1}R))^{\prime
}-\mu _{2}k_{1}^{2}j_{\nu}(k_{1}R)(k_{2}Rh_{\nu }^{1}(k_{2}R))^{\prime }%
\right] \biggr\},  \notag  \label{eq:det}
\end{eqnarray}
and the regularized Casimir energy can be written as \cite{bomu01,vanka68}: 
\begin{equation}
E_{reg}=\frac{1}{2\pi }\sum_{l=1}^{\infty }(2l+1)\int_{0}^{\infty }{d\omega
\ln \frac{Det(i\omega ,R)}{Det(i\omega ,R\rightarrow \infty )}}
\end{equation}

Along the imaginary axis we have 
\begin{eqnarray}
Det(i\omega ,R) &=&\frac{\theta ^{2}}{\epsilon _{1}\mu _{1}\epsilon
_{2}\mu_{2}\omega ^{4}R^{2}}[e_{\nu }(k_{2}R)s_{\nu }(k_{1}R)e_{\nu
}^{\prime}(k_{2}R)s_{\nu }^{\prime }(k_{1}R)]+ \\
&&\frac{\epsilon _{1}\epsilon _{2}}{(\epsilon _{1}\mu
_{1})^{2}(\epsilon_{2}\mu _{2})^{2}\omega ^{4}R^{2}}\left[ \sqrt{\epsilon
_{2}\mu _{1}}s_{\nu}^{\prime }(k_{1}R)e_{\nu }(k_{2}R)-\sqrt{\epsilon
_{1}\mu _{2}}e^{\prime}(k_{2}R)s_{\nu }(k_{1}R)\right] \times  \notag \\
&&\left[ \sqrt{\epsilon _{2}\mu _{1}}s_{\nu }(k_{1}R)e_{\nu
}^{\prime}(k_{2}R)-\sqrt{\epsilon _{1}\mu _{2}}e(k_{2}R)s_{\nu }^{\prime
}(k_{1}R) \right] .  \notag  \label{eq:deti}
\end{eqnarray}
with $e_{\nu}(x)=\sqrt{\frac{2x}{\pi}}K_{l+1/2}$ and $s_{\nu}(x)=\sqrt{\frac{%
\pi x}{2}}I_{l+1/2}$, and $I_{\nu}(x)=i^{-\nu}J_{\nu}(ix)$, $%
K_{\nu}(x)=i^{\nu +1}\frac{\pi}{2}H_{\nu}^1(ix)$.

Note that for $\theta=0$ the usual result \cite{bomu01} is recovered. Assuming equal speed of light inside and outside, $\epsilon_1\mu_1=\epsilon_2\mu_2=1$, the expression greatly simplifies and we get (using the expression for the wronskian $s^{\prime }(x)e(x)-e^{\prime }(x)s(x)=1$),

\begin{equation}
Det(i\omega,R) =- \theta^2\frac{[1-\left[(s_\nu(\omega R)e_\nu(\omega
R))^{\prime }\right]^2]}{4\omega^4 R^2}- \frac{[1-\xi^2\left[(s_\nu(\omega
R)e_\nu(\omega R))^{\prime }\right]^2]}{\frac{(1+\xi)^2}{\epsilon_1^2}%
\omega^4 R^2}
\end{equation}
with $\xi=\frac{\epsilon_1-\epsilon_2}{\epsilon_1+\epsilon_2}$.

Using the asymptotic expression for $x\rightarrow \infty $ of $I_{n}$ and $%
K_{n}$: $I_{n}(x)\simeq \frac{e^{x}}{\sqrt{2\pi x}}$, $K_{n}(x)\simeq e^{-x}%
\sqrt{\frac{\pi }{2x}}$ \cite{abste}, one finds 
\begin{equation}
Det(i\omega, R\rightarrow \infty)\simeq -\frac{1}{4\omega^4R^2}\left[%
\theta^2+\frac{4\epsilon_1^2}{(1+\xi )^2}\right] \,.
\end{equation}
Defining $\lambda _{\nu }(y)=(s_{\nu }(y)e_{\nu }(y))^{\prime }$, the
regularized Casimir energy is 
\begin{eqnarray}
E_{reg} &=&\frac{1}{\pi R}\sum_{l=1}^{\infty }(l+1/2)\int_{0}^{\infty }{\ dy
\ln \left\{ \frac{4[1-\xi ^{2}\lambda _{\nu }^{2}(y)]+\theta ^{2}\frac{
(1+\xi )^{2}}{\epsilon _{1}^{2}}[1-\lambda _{\nu }^{2}(y)]}{\theta ^{2}\frac{
(1+\xi )^{2}}{\epsilon _{1}^{2}}+4}\right\} }  \notag \\
&=&\frac{1}{\pi R}\sum_{l=1}^{\infty }(l+1/2)\int_{0}^{\infty }{\ dy\ln
[1-\eta \lambda _{\nu }^{2}(y)]}  \label{casi2'}
\end{eqnarray}
with $y=\omega R$ and $\eta =\frac{4\xi^2 \epsilon_1^2+\theta^2(1+\xi )^2}{%
4\epsilon_1^2+\theta^2(1+\xi )^2}= \frac{(\epsilon_1-\epsilon_2)^2 + \theta^2%
}{(\epsilon_1+\epsilon_2)^2+\theta^2}\leq 1$.

We observe that the $\theta$ contribution simply reduces to a change of parameters: $\xi\rightarrow\eta$. The integral is convergent see \cite{nespi98}, but the $\nu=(l+1/2)$ series need to be regularized. This can be done by different methods \cite{milton78,nespi98}, probably the easiest is to use $\zeta$ function. To this end we define 
\begin{eqnarray*}
E &=& \lim_{s\rightarrow0}E_{reg}\nu^{-s}= \frac{1}{\pi R}
\sum_{l=1}^\infty(l+1/2)^{(2-s)}\int{\ dz\ln [1-\eta \lambda_\nu^2(\nu z)] }
\\
&\simeq& \lim_{s\rightarrow0} \frac{1}{\pi R}\sum_{l=1}^\infty\nu^{(2-s)}
\int {\ dz\ln [1-\frac{\eta}{\nu^24(1+z^2)^3}] }
\end{eqnarray*}
where $y=\nu z$, and the leading order asymptotic expansion with respect $%
\nu\rightarrow\infty$: $s_{\nu}(\nu z)e_{\nu}(\nu z)\sim \frac{z}{2\sqrt{%
1+z^2}}$ has been used \cite{abste}. To sum the series we need the
expression of $E$ for large values of $\nu$: 
\begin{equation}
E=\frac{1}{\pi R}\sum_{l=1}^\infty\nu^{(2-s)}\int{\ dz \left[-\frac{\eta}{%
4(1+z^2)^3\nu^2}-\frac{36\eta^2}{1152(1+z^2)^6\nu^4}+O(\nu^{-6}) \right] } .
\end{equation}
Thus, 
\begin{eqnarray}
E &=&\lim_{s\rightarrow0} \frac{1}{\pi R}\sum_{l=1}^\infty {\ \left[-\frac{%
3\pi\eta}{64}\frac{1}{(l+\frac{1}{2})^s}-\frac{756\pi\eta^2}{ 196608} \frac{%
1 }{(l+\frac{1}{2})^{(2+s)}} \right] }  \notag \\
&=& \lim_{s\rightarrow0}\left\{-\frac{3\eta}{64R}\zeta(s,3/2)+\frac{27\eta}{%
8192R}\zeta(2+s,3/2)-\frac{63\eta^2}{16384R}\zeta(2+s,3/2)\right\}  \notag \\
&=&\frac{3\eta}{64R}-\frac{27\eta}{16384R}(\pi^2-8)+\frac{63\eta^2}{32768R}
(\pi^2-8).  \label{eq:ene}
\end{eqnarray}
where $\zeta(s,p)$ is the Hurwitz zeta function.\newline
------------------------ \newline
\textbf{{\large {Acknowledgements}}} \newline We warmly thank Steve Willison for enlightening discussions on the Atiyah-Patodi-Singer theorem. This work is partially supported by FONDECYT grants 11080056, 1100755, 1085322, 1100328, and 1110102, and by the ``Southern Theoretical Physics Laboratory'' ACT-91 grant from CONICYT. The Centro de Estudios Cient\'{\i}ficos (CECs) is funded by the Chilean Government through the Centers of Excellence Base Financing Program of CONICYT. F. C. is also supported by Proyecto de Inserci\'on CONICYT 79090034, and by the Agenzia Spaziale Italiana (ASI).

\end{document}